\documentclass[showpacs,pre,aps,twocolumn,superscriptaddress]{revtex4-1}

\pdfoutput=1
\usepackage{amsmath,amsfonts,amssymb,bm,graphicx,hyperref,color,units}

\newcommand{\llangle}{\langle\!\langle}
\newcommand{\rrangle}{\rangle\!\rangle}
\newcommand{\bra}[1]{\langle #1|}
\newcommand{\ket}[1]{|#1\rangle}
\newcommand{\Tr} {{\text T}{\text r}}
\newcommand{\revision}[1]{{#1}}
\newcommand{\revisiontwo}[1]{{#1}}

\begin{document}

\title{Intermittency and dynamical Lee-Yang zeros of open quantum systems}
\author{James M. Hickey}
\affiliation{School of Physics and Astronomy, University of Nottingham, Nottingham, NG7 2RD, United Kingdom}
\author{Christian Flindt}
\affiliation{D\'epartement de Physique Th\'eorique, Universit\'e de Gen\`eve, 1211 Gen\`eve, Switzerland}
\author{Juan P. Garrahan}
\affiliation{School of Physics and Astronomy, University of Nottingham, Nottingham, NG7 2RD, United Kingdom}

\date{\today}



\begin{abstract}
We use high-order cumulants to investigate the Lee-Yang zeros of generating functions of dynamical observables in open quantum systems. At long times the generating functions take on a large deviation form with singularities of the associated cumulant generating functions --- or dynamical free energies --- signifying phase transitions in the ensemble of dynamical trajectories.  We consider a driven three-level system as well as the dissipative Ising model. Both systems exhibit dynamical intermittency in the statistics of quantum jumps. From the short-time behavior of the dynamical Lee-Yang zeros we identify critical values of the counting field which we attribute to the observed intermittency and dynamical phase co-existence.  Furthermore, for the dissipative Ising model we construct a trajectory phase diagram and estimate the value of the transverse field where the stationary state changes from being ferromagnetic (inactive) to being paramagnetic (active).
\end{abstract}

\maketitle

\section{Introduction}
\label{sec:Intro}

In many-body systems, both classical and quantum, dynamics is often more than statics: collective effects make the dynamical behavior much richer than what would be expected based on the static properties alone.  A proper characterization of such complex emergent dynamics requires tools that directly address the statistical properties of dynamical trajectories.  One such set of tools can be fashioned from the large-deviation (LD) method \cite{Demboo1998,Touchette2009}.  By considering the LD properties of time-integrated observables one can construct a ``thermodynamics of trajectories'', i.~e., a dynamical equivalent to the equilibrium ensemble method of statistical mechanics.  This is essentially Ruelle's thermodynamic formalism for dynamical systems \cite{Eckmann1985,Ruelle2004} adapted to stochastic dynamics \cite{Merolle2005,Lecomte2007}. This approach, sometimes called the $s$-ensemble formalism \cite{Hedges2009}, has successfully been applied to a variety of classical stochastic problems \cite{Merolle2005,Lecomte2007,Garrahan2007,Hedges2009,Giardina2006,Baule2008,Gorissen2009,Jack2010,Giardina2011,Hurtado2011,Nemoto2011,Bodineau2012a,Bodineau2012b,Chetrite2013,Chetrite2014}.

The $s$-ensemble formalism can also be applied to open quantum systems \cite{Garrahan2010}.  In this case, time-extensive observables can be the total number of photon emissions~\cite{Cook1981,Lenstra1982}, the number of electrons transported through a sub-micron conductor~\cite{Levitov1993,Levitov1996,Pilgram2003,Flindt2008} or the time-integrated quadratures of the emitted light~\cite{Hickey2012}. The generating functions of such observables can be obtained using full counting statistics (FCS) techniques which emerged in parallel from the fields of electronic transport~\cite{Nazarov2003,Esposito2009}, quantum optics~\cite{Plenio1998,Gardiner2004} and classical
stochastic processes~\cite{Gardiner1986,VanKampen2007}.

Recently, we have shown~\cite{Flindt2013} that the analogy between the $s$-ensemble method and standard equilibrium ensembles extends also to a dynamical generalization of the Lee-Yang theory of equilibrium phase transitions~\cite{Lee1952,Yang1952,Blythe2002,Bena2005,Wei2012}, which connects complex singularities of partition sums in finite systems with equilibrium phase transitions in the thermodynamic limit. Specifically, we showed that the complex zeros of the moment generating function (MGF) of a dynamical observable of interest are directly connected with the short-time behavior of the high-order cumulants of the observable~\cite{Flindt2009,Flindt2010,Kambly2011}.  We demonstrated that one may determine the dynamical Lee-Yang zeros from the high-order cumulants, and from the time-evolution of these zeros infer the existence and location of dynamical phase transitions. This method, which was applied to classical stochastic many-body systems, specifically kinetically constrained models of glassy systems~\cite{Ritort2003} and the one-dimensional Glauber-Ising chain~\cite{Hickey2013E}, shows that singularities associated with counting fields \cite{Levkivskyi2009,Ivanov2010,Li2011,Ivanov2013,Utsumi2013,Ren2013} are readily accessible from physical observables.

\begin{figure*}
\includegraphics[width = 0.92\textwidth]{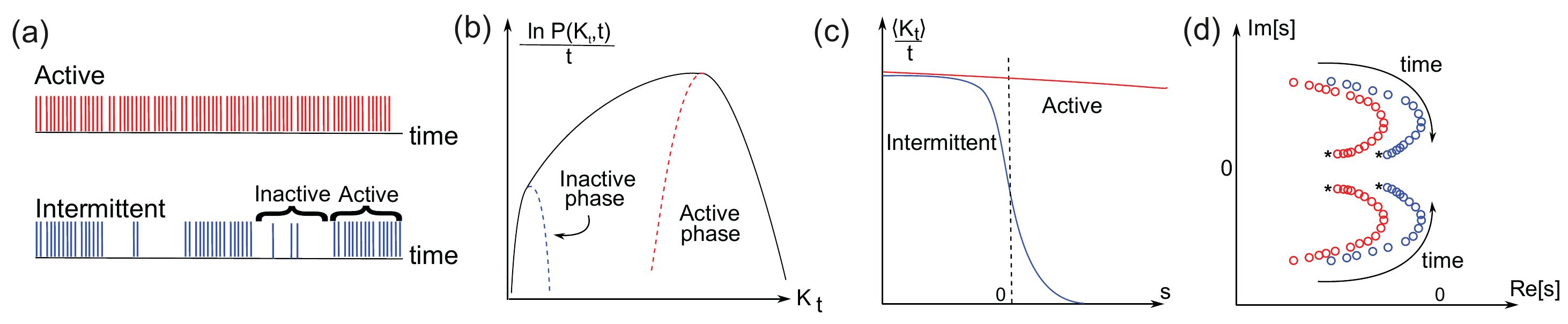}
\caption{(Color online) Thermodynamics of trajectories and the $s$-ensemble formalism. (a) Time-traces of quantum jumps. In the active phase, a large number of quantum jumps occur per unit time. When the system is intermittent, it switches randomly between the active phase and the inactive phase. (b) Large deviation function $(1/t)\ln P(K_t,t)$ at long times. The tails of the distribution are determined by the fluctuations in each of the two phases. The intermediate region is governed by the random switching between the two phases. (c) Dynamical activity (number of quantum jumps) per unit time as a function of the counting field $s$. By tuning the counting field, the system may be driven through a phase transition (here close to $s=0$), where it changes from being in an active phase with a large dynamical activity to an inactive phase with a low dynamical activity. Here we show a smeared first-order transition. (d) Dynamical Lee-Yang zeros in the complex plane of the counting field. \revision{With time}, the dynamical Lee-Yang zeros move towards the complex values of the counting field, where the dynamical free energy (the CGF) displays singular features corresponding to a phase transition. The phase transition is smeared when it occurs for complex values of the counting field, and it only becomes sharp as the transition points move on to the real-axis.}
\label{fig:fig1}
\end{figure*}

In this work we extend our technique to the realm of open quantum systems with specific focus on the dynamical phenomenon of intermittency. This behavior of intervals of high dynamical activity (for instance a large number of photon emissions), interspersed with intervals of low dynamical activity, see Fig.~\ref{fig:fig1}a, has been explained in terms of dynamical phase coexistence~\cite{Garrahan2010,Ates2012}. The intermittent behavior is reflected in the LD function, see Fig.~\ref{fig:fig1}b, and the cumulant generating function (CGF) of the dynamical activity has (smoothed) first-order singularities associated with the two dynamical phases, see Fig.~\ref{fig:fig1}c. As we have recently demonstrated~\cite{Flindt2013}, the location of these singularities  can be deduced from the high-order cumulants and the dynamical Lee-Yang zeros which with time move towards the singular points in the complex plane of the counting field, see Fig.~\ref{fig:fig1}d.

Below we focus on two systems which exhibit dynamical intermittency. We first consider a simple driven quantum three-level system before moving on to the dissipative Ising chain as an example of a quantum many-body system.  Both systems exhibit dynamical crossovers which result in dynamical intermittency in the real-time dynamics~\cite{Garrahan2010,Ates2012}. Treating the counting field as a complex variable we demonstrate that these systems have trajectory critical points which appear as complex conjugate pairs in the complex plane of the counting field. \revision{These phase transitions do not have to occur at zero counting field corresponding to the actual physical dynamics.} As we tune the system parameters to make the crossovers sharp, the critical points move towards the origin in the complex plane, where the physical dynamics takes place. For the dissipative Ising chain, where the crossover becomes a sharp transition in the limit of an infinitely long chain, we show that one may estimate the positions of the stationary state transition points without resorting to mean-field methods and we construct the associated trajectory phase diagram. Our work shows that one may use short-time cumulants to infer the positions of trajectory critical points in open quantum systems at nonzero values of the counting field.

\revision{Our work establishes firmly that our method, based on dynamical Lee-Yang zeros and originally developed for classical stochastic processes, can be extended to open quantum systems. There is currently much interest in the non-equilibrium dynamics of open quantum systems and our work contributes to this field by extending our method for classical non-equilibrium systems to this area of quantum dynamics. For quantum systems, this is particularly important given the usual complications in their numerical simulation which preclude, to date, the use of advanced methods to sample rare fluctuations which are readily available for classical systems: as we emphasize in the paper, our method allows us to probe singularities in the dynamical generating functions through the standard unbiased dynamics of the system (rather than through biased trajectory ensembles, as for example with classical transition path sampling methods).}

The paper is now organized as follows. In Sec.~\ref{sec:Thermo} we introduce the $s$-ensemble formalism.  In Sec.~\ref{sec:LYZ} we describe our method based on high-order cumulants and dynamical Lee-Yang zeros \cite{Flindt2013}. The driven three-level system, along with its results, is discussed in Sec.~\ref{sec:3level}.  In Sec.~\ref{sec:Ising} we move on to the dissipative Ising model for which we construct a trajectory phase diagram. Finally, in Sec.~\ref{sec:Conc} we present our conclusions.

\section{Thermodynamics of Trajectories}
\label{sec:Thermo}

We examine the statistics of dynamical trajectories in open quantum systems using a thermodynamic formalism known as the $s$-ensemble. Within this approach, the dynamical trajectories play the role of microstates in statistical mechanics following Ruelle \cite{Ruelle2004} (see also Refs.~\cite{Eckmann1985,Gaspard2005,Touchette2009} for comprehensive reviews). The trajectories are classified according to an associated time-extensive quantity $K_t$ which we refer to as the \emph{dynamical activity}. The dynamical activity is assumed to be a (positive) integer variable, for instance the number of electrons that have been transferred through an electrical conductor ~\cite{Levitov1993,Levitov1996,Pilgram2003,Flindt2008}, the number of photons that have been emitted from a light source~\cite{Cook1981,Lenstra1982}, or, more generally, the number of quantum jumps that have occurred in an open quantum system during the time span $[0,t]$. Continuous variables can also be treated with only minor modifications, for example the total dissipated heat or the time-integrated energy as in Ref.~\cite{Hickey2013E}.

We focus on open quantum systems described by a reduced density matrix $\hat{\rho}(t)$ whose dynamics is governed by a generalized master equation (GME) of the form~\cite{Gardiner2004,Gorini1976,Lindblad1976}
\begin{equation}
\label{eq:master}
\begin{split}
\frac{d}{dt}\hat{\rho}(t) &= \mathcal{L}\hat{\rho}(t)\\
&=-i[\hat{H}, \hat{\rho}(t)] + \sum_{j} \left[\hat{L}_{j}\hat{\rho}(t) \hat{L}^{\dagger}_{j} -
\frac{1}{2} \left\{ \hat{L}^{\dagger}_{j}\hat{L}_{j},\hat{\rho}(t) \right\}\right].
\end{split}
\end{equation}
The Liouvillian $\mathcal{L}$ consists of two parts: The commutator describes the coherent evolution of the system itself governed by the Hamiltonian $\hat{H}$.  The Lindblad operators $\hat{L}_{j}$ describe incoherent quantum jumps due to interactions with the environment. We consider here Markovian generalized master equations, but an extension to non-Markovian systems is also possible, see e.~g.~\cite{Flindt2008,Flindt2010}.

We are interested in the statistics of quantum jumps that have occurred during the time span $[0,t]$. We do not discriminate between the different types of quantum jumps corresponding to each Lindblad operator $\hat{L}_{j}$ although such a distinction can easily be implemented. We proceed by resolving the density matrix with respect to the number of quantum jumps $K_t$ that have occurred and denote this $K_t$-resolved density matrix as $\hat{\rho}(K_t,t)$~\cite{Plenio1998,Makhlin2001}. From the $K_t$-resolved density matrix we obtain the distribution of the dynamical activity by tracing out the system degrees of freedom, i.~e.
\begin{equation}
P(K_t,t) =  \Tr[\hat{\rho}(K_t,t)].
\end{equation}
Furthermore, we may define a MGF corresponding to $P(K_t,t)$ as
\begin{equation}
\label{eq:MGFmas}
Z(s,t) = \sum_{K_t} {e}^{-s K_t} P(K_t,t),
\end{equation}
where $s$ is referred to as the counting field. The MGF can conveniently be expressed as
\begin{equation}
\label{eq:MGFmas2}
Z(s,t) = \Tr[\hat{\rho}(s,t)]
\end{equation}
in terms of the $s$-dependent density matrix
\begin{equation}
\hat{\rho}(s,t) = \sum_{K_t}{e}^{-s K_t} \hat{\rho}(K_t,t).
\end{equation}
We note that by setting $s=0$, we recover the original density matrix,
\begin{equation}
\hat{\rho}(s=0,t)=\hat{\rho}(t).
\end{equation}
In addition, by differentiating the MGF with respect to the counting field evaluated at $s = 0$, we obtain the moments of $K_t$ as
\begin{equation}
\langle K_t^{n}\rangle = (-1)^{n}\partial^{n}_{s}Z(s,t)|_{s\rightarrow0}.
\end{equation}
We also define the CGF
\begin{equation}
\label{eq:CGFmas}
\Theta(s,t)= \ln{Z(s,t)}
\end{equation}
from which the cumulants of $K_t$ follow by differentiation with respect to the counting field at $s = 0$
\begin{equation}
\llangle K_t^{n}\rrangle =
(-1)^{n}\partial^{n}_{s}\Theta(s,t)|_{s\rightarrow0}.
\end{equation}

To evaluate the MGF we consider a modified GME governing the evolution of the $s$-dependent density matrix. The modified GME takes the form
\begin{equation}
\label{eq:smaster}
\frac{d}{dt}\hat{\rho}(s,t) = \mathcal{L}(s)\hat{\rho}(s,t),
\end{equation}
where the $s$-dependent Liouvillian $\mathcal{L}(s)$ acts on density matrices as
\begin{equation}
\mathcal{L}(s)\hat{\rho}(s,t)= \mathcal{L}\hat{\rho}(s,t)+ \sum_{j} ({e}^{-s}-1)\hat{L}_{j}\hat{\rho}(s,t) \hat{L}^{\dagger}_{j}.
\label{eq:sliouv}
\end{equation}
Here the terms multiplied by $(e^{-s}-1)$ increase the number of quantum jumps by one, i.~e.,~$K_t\rightarrow K_t+1$. Since Eq.~(\ref{eq:smaster}) is a linear differential equation for $\hat{\rho}(s,t)$, it can be rewritten in the form
\begin{equation}
\label{eq:matrixrep}
\frac{d}{dt}\ket{\rho(s,t)}\!\rangle= \mathbb{L}(s)\ket{\rho(s,t)}\!\rangle,
\end{equation}
where $\ket{\rho(s,t)}\!\rangle$ is the vector representation of $\hat{\rho}(s,t)$ in a suitable basis and $\mathbb{L}(s)$ is the matrix representation of $\mathcal{L}(s)$. (Here we use double angle brackets to distinguish these vectors from the ordinary quantum mechanical `bras' and `kets' used later on.) Furthermore, formally solving Eq.~(\ref{eq:matrixrep}) we find
\begin{equation}
\label{eq:matrixsol}
\ket{\rho(s,t)}\!\rangle= e^{\mathbb{L}(s)t}\ket{0}\!\rangle,
\end{equation}
assuming that the system at $t=0$ has reached the stationary state defined by $\mathbb{L}(0)\ket{0}\!\rangle=0$. Since the Liouvillian $\mathcal{L}$ in Eq.~(\ref{eq:master}) conserves probability, it holds that $\Tr[\mathcal{L}\hat{\rho}(t)]=0$ for any density matrix. This implies that the left zero-eigenvector of $\mathbb{L}(0)$ is the vector representation of the trace operation, i.~e., $\langle\!\bra{\tilde{0}}\mathbb{L}(0)=0$ and $\langle\!\langle\tilde{0}|{0}\rangle\!\rangle=\Tr[\hat{\rho}(s=0,t)]=1$. Combining Eqs.~(\ref{eq:MGFmas2}) and (\ref{eq:matrixsol}) we then obtain the following compact expression
\begin{equation}
\label{eq:MGFcomp}
Z(s,t)= \langle\!\bra{\tilde{0}}e^{\mathbb{L}(s)t}\ket{0}\!\rangle.
\end{equation}
\revision{This expression holds for a system that has been prepared in an arbitrary state in the far past. The systems has then evolved until $t=0$, where it has reached the stationary state. At $t=0$, we start collecting statistics to construct the probability distribution $P(K_t,t)$ for the number of quantum jumps that have occurred in the time span $[0, t]$. Now, from Eq.~(\ref{eq:MGFcomp}) we} see that the MGF and the CGF at long times take on the large deviation forms
\begin{equation}
Z(s,t) \simeq {e}^{t\theta(s)}
\end{equation}
and
\begin{equation}
\Theta(s,t) \simeq t\theta(s),
\end{equation}
where
\begin{equation}
\theta(s) = \max_{j}[\lambda_{j}(s)]
\label{eq:maxlambda}
\end{equation}
is the eigenvalue in the spectrum $\{\lambda_{j}(s)\}$ of $\mathbb{L}(s)$ with the largest real-part.

Having established this general framework, we are now ready to formulate the basic principles of the $s$-ensemble formalism. Drawing on the analogy between dynamical trajectories in open quantum systems and the microstates in statistical mechanics, we consider the MGF as a (dynamical) partition function, the CGF as a (dynamical) free energy, and the counting field $s$ as an external field that biases the ensemble of trajectories away from the typical dynamics at $s = 0$. In addition, time plays the extensive role of volume with the limit of long times corresponding to the thermodynamic limit in statistical mechanics. Importantly, by changing the counting field $s$ one may drive the system across a phase transition between different dynamical phases~\cite{Garrahan2010,Garrahan2011,Genway2012,Hickey2012}.  Analogously to
equilibrium statistical mechanics such trajectory phase transitions are manifested as singular features in the dynamical free energy~$\theta(s)$.

The counting field $s$ can be treated as a complex parameter. \revision{At $s=0$ the modified GME in Eq.~(\ref{eq:smaster}) returns to the Lindblad form in Eq.~\eqref{eq:master}, and the Liouvillian has a single zero eigenvalue $\lambda_{0}(s = 0) = 0$ corresponding to the stationary state.  All other eigenvalues have negative real-parts which ensure exponential relaxation towards the stationary state. However, with a non-zero counting field the two largest eigenvalues may cross each other giving rise to singular features in the dynamical free energy~$\theta(s)$.} The singular behaviors in the dynamical free energy~$\theta(s)$ occur at critical values of the counting field $s=s_c$, where the two largest eigenvalues of $\mathbb{L}(s)$ are degenerate, i.~e.
\begin{equation}
\lambda_{0}(s_c)=\lambda_{1}(s_c).
\label{eq:phasetrans}
\end{equation}
\revision{As the two largest eigenvalues cross, the dynamical free energy~$\theta(s)$ in Eq.~(\ref{eq:maxlambda}) may display a kink, such that the first derivative (the dynamical activity, see Fig.~\ref{fig:fig1}c) or higher derivatives are discontinuous, signaling a phase transition.} To characterize the trajectory phase transitions, one may consider the long-time limit of the average dynamical activity per unit time as a function of $s$
\begin{equation}
k(s) = \lim_{t\rightarrow\infty} \frac{\langle K_t \rangle(s)}{t} = -\partial_{s} \theta(s).
\label{eq:ave_dyn_act}
\end{equation}
A discontinuity in the average activity is then indicative of a first-order trajectory phase transition. Similarly, a discontinuity in the higher derivatives of the dynamical free energy corresponds to a continuous phase transition.

It is clear from the definitions in Eqs.~(\ref{eq:MGFmas}) and (\ref{eq:CGFmas}) that
\begin{equation}
Z(s^{*},t)= Z^{*}(s,t)
\end{equation}
and
\begin{equation}
\Theta(s^{*},t) = \Theta^{*}(s,t).
\label{eq:theta_conj}
\end{equation}
Moreover, if $\theta(s)$ is singular at $s=s_c$, then so is $\theta^*(s)$. Using the equations above, we then find that $s^*_c$ must also be a transition point of $\theta(s)$. Complex trajectory transition points thus appear as complex conjugate pairs. Furthermore, the MGF is $2\pi$-periodic along the imaginary axis, implying that all the points $s_c + i2\pi n$ with $n \in \mathbb{Z}$ are transition points of $\theta(s)$.  However, it suffices to find the complex conjugate pair of transition points in the strip  $-\pi<\mathrm{Im}(s)\leq\pi$ from which the rest follows.

The $s$-ensemble formalism has found use in several contexts. It has for instance been suggested that a trajectory phase transition underlies the glass transition in liquids~\cite{Garrahan2007,Hedges2009,Pitard2011,Speck2012,Bodineau2012a}. The formalism has also been used to understand the thermodynamics of quantum jump trajectories \cite{Garrahan2010}, for example in micromasers \cite{Garrahan2011} and superconducting single-electron transistors coupled to a resonator \cite{Genway2012}. In addition, the formalism has been applied in investigations of quadrature trajectories \cite{Hickey2012} as well as dynamical crossovers in exciton transport \cite{Hossein-Nejad2013}. Recently, we examined trajectory phase transitions in the Glauber-Ising model which exhibits a whole curve of critical points in the complex plane of the counting field \cite{Jack2010,Hickey2013E}.

However, despite its usefulness as a theoretical tool, an experimental realization of the $s$-ensemble formalism faces two apparent problems: Firstly, trajectory phase transitions may occur at non-zero values of the counting field ($s\neq0$), making them difficult to observe, since the real dynamics takes place without the counting field ($s=0$). Secondly, the singularities in the dynamical free energy only appear in the limit of long times, which in practice may be out of reach \revisiontwo{as
the probabilities decay exponentially with time and their values become extremely small.} To resolve these problems, we suggested in Ref.~\cite{Flindt2013} to employ a dynamical generalization of ideas from equilibrium statistical mechanics by Yang and Lee by connecting the motion of the zeros of the MGF with the short-time behavior of the high-order cumulants of the dynamical activity.

Specifically, we showed that the positions of the dynamical Lee-Yang zeros can be inferred from the high-order cumulants of the dynamical activity, making it possible to follow the zeros as they with time move towards the location in the complex plane of the counting field, where the trajectory phase transitions occur. Using this approach, we showed how the finite-time behavior of the high-order cumulants can be used to infer the existence and location of trajectory phase transitions in stochastic many-body systems as for instance facilitated spin models of glasses and the Glauber-Ising model in Ref.~\cite{Hickey2013E}.  \revision{Importantly, the extraction of the dynamical Lee-Yang zeros relies only on the high-order cumulants obtained with the counting field set to zero ($s = 0$), and those are in principle directly accessible in an experiment.}

For the sake of completeness, we now provide the essential elements of the method developed in Ref.~\cite{Flindt2013} \revision{and demonstrate that it can readily be applied also in the context of open quantum systems}.

\section{Dynamical Lee-Yang zeros}
\label{sec:LYZ}

We follow the ideas of Lee and Yang by considering the (dynamical) zeros $s_j(t)$ of the MGF as functions of time. Close to a transition point $s=s_c$, where the two largest eigenvalues are nearly degenerate, $\lambda_{0}(s)\simeq \lambda_{1}(s)$, we may approximate the MGF as
\begin{equation}
Z(s,t)  \simeq c_0(s)e^{\lambda_0(s)t}+c_1(s)e^{\lambda_1(s)t}
\end{equation}
by neglecting the contributions from the other eigenvalues with smaller real-parts. The prefactors $c_{0}(s)$ and $c_{1}(s)$ are determined by the initial conditions at $t=0$ and are not important in the following. From this approximation, we readily find that the zeros of the MGF are given by the equations
\begin{equation}
\lambda_0(s)=\lambda_1(s)+\frac{\ln[c_1(s)/c_0(s)]+i\pi(2n+1)}{t},
\end{equation}
where $n$ is an integer. At long times, the second term on the right-hand side becomes arbitrarily small, such that the equations for the Lee-Yang zeros eventually reduce to $\lambda_{0}(s) = \lambda_{1}(s)$; see also Eq.~(\ref{eq:phasetrans}). This shows us that the dynamical Lee-Yang zeros of the MGF will move to the transition point at $s = s_{c}$ as time increases.

Having established this connection, we need to relate the Lee-Yang zeros with the measurable fluctuations encoded in the cumulants of $K_t$. To this end we factorize the MGF in terms of the Lee-Yang zeros as
\begin{equation}
Z(s,t) \propto \prod_{j}\left[\frac{s_{j}(t)-s}{s_{j}(t)}\right],
\end{equation}
having omitted any analytic prefactors. \revision{In principle, there can be infinitely many Lee-Yang zeros which we denote as $s_j(t)$}. We note that the MGF at finite times is real and positive for real values of $s$, such that the dynamical Lee-Yang zeros must be non-real and come in complex conjugate pairs in accordance with the discussion following Eq.~(\ref{eq:theta_conj}). We then find for the associated CGF
\begin{equation}
\label{eq:LY}
\Theta(s,t)\simeq\sum_{j}\left(\ln[s_{j}(t)-s]-\ln[s_{j}(t)]\right),
\end{equation}
again having omitted any analytic terms in the CGF. We see that the Lee-Yang zeros appear as (logarithmic) singularities of the CGF, which determine its high derivatives according to Darboux's theorem~\cite{Dingle1973,Berry2005}. Indeed, by differentiating the CGF with respect to $s$, we find that the high-order cumulants of $K_{t}$ are given as~\cite{Flindt2009,Flindt2010,Kambly2011}
\begin{equation}
\llangle K_t^{n}\rrangle \simeq
(-1)^{(n-1)}(n-1)!\sum_{j}\frac{{e}^{-in\arg[{s}_{j}(t)]}}{|s_{j}(t)|^{n}},
\label{eq:highcumu}
\end{equation}
where
\begin{equation}
s_{j}(t) = |s_{j}(t)|{e}^{i \arg[s_{j}]}.
\end{equation}
Moreover, for large orders $n$, the sum in Eq.~(\ref{eq:highcumu}) is dominated by the pair of Lee-Yang zeros closest to the
origin, which we denote as $s_{0}(t)$ and $s^{*}_{0}(t)$.  We may then further approximate the sum as~\cite{Bhalerao2003,Berry2005,Flindt2009,Flindt2010,Kambly2011,Flindt2013}
\begin{equation}
\label{eq:approx}
\llangle K_t^{n}\rrangle\simeq(-1)^{(n-1)}(n-1)!\frac{2\cos[n\arg
s_{0}(t)]}{|s_{0}(t)|^{n}}.
\end{equation}
\revision{This result is attractive as it allows us to determine the leading pair of Lee-Yang zeros $s_{0}(t)$ and $s^{*}_{0}(t)$ from the high-order cumulants of the dynamical activity obtained with $s=0$. Equation~(\ref{eq:approx}) makes it possible to extract only the leading pair of Lee-Yang zeros closest to $s=0$ and discard all other Lee-Yang zeros. The approximation in Eq.~(\ref{eq:approx}) improves as the order $n$ is increased and the contributions from the other zeros are suppressed.}

Next, the following matrix equation can be derived from Eq.~(\ref{eq:approx})~\cite{Zamastil2005,Flindt2010,Kambly2011,Flindt2013}
\begin{equation}
\label{eq:MatEq}
\left[
\begin{array}{cc}
1 & -\frac{\kappa^{(+)}_{n}}{n} \vspace{.05cm} \\
1 & -\frac{\kappa^{(+)}_{n+1}}{n+1}\\
\end{array}
\right]
\cdot
\left[
\begin{array}{c}
-(s_{0}+s^{*}_{0}) \\
|s_{0}|^{2}\\
\end{array}
\right]
=
\left[
\begin{array}{c}
(n-1)\kappa^{(-)}_{n}\vspace{.25cm}\\
n\kappa^{(-)}_{n+1}\\
\end{array}
\right],
\end{equation}
given the ratios of cumulants
\begin{equation}
\kappa^{(\pm)}_{n}(t)\equiv \frac{\llangle
K_t^{n\pm1}\rrangle}{\llangle K_t^{n}\rrangle}.
\end{equation}
By solving this matrix equation, we may extract the leading pair of dynamical Lee-Yang zeros closest to the origin, $s_{0}(t)$ and $s^{*}_{0}(t)$, from four consecutive cumulants.

\revision{In Fig.~\ref{fig:fig1}c we show a smeared crossover as the counting field $s$ is tuned from negative real values to positive real values. Such a smooth crossover is indicative of an avoided crossing between the two largest eigenvalues for real values of the counting field. The sharpness of the crossover is determined by the dynamical susceptibility
\begin{equation}
\theta''(s)=\partial^2_s\theta(s).
\end{equation}
In a simple picture, we may evaluate the susceptibility by factorizing the MGF as
\begin{equation}
Z(s,t) \simeq \prod_{j}\frac{{e}^{-s_j(t)}-{e}^{-s}}{e^{-s_j(t)}},
\end{equation}
recalling that the Lee-Yang zeros are denoted as $s_j(t)$. We then approximate the dynamical free energy as
\begin{equation}
\begin{split}
\theta(s) &= \lim_{t\rightarrow \infty}\frac{\ln Z(s,t)}{t}\\
&\simeq\ln\left(\frac{{e}^{-s_c}-{e}^{-s}}{e^{-s_c}}\right)+\ln\left(\frac{{e}^{-s_c^*}-{e}^{-s}}{e^{-s_c^*}}\right)
\end{split}
\end{equation}
by simply replacing the Lee-Yang zeros by the complex pair of critical points $s_c$ and $s_c^*$, assuming that the prefactor $1/t$ is compensated by the diverging density of zeros at the critical points in the limit of long times. Using this approximation we readily find
\begin{equation}
\theta''(s) \simeq -\frac{1}{4}\left(\frac{1}{\sinh^2[(s-s_c)/2]}+\frac{1}{\sinh^2[(s-s_c^*)/2]}\right).
\nonumber
\end{equation}
Evaluating the dynamical susceptibility at the real part of the critical points, $s=\mathrm{Re}[s_c]$, we obtain
\begin{equation}
\label{eq:Relate1}
\theta''(s=\mathrm{Re}[s_c]) \simeq \frac{1}{2\sin^2(\mathrm{Im}[s_c]/2)} \simeq \frac{2}{\mathrm{Im}[s_c]^2},
\end{equation}
assuming that the critical points are close to the real-axis, $\mathrm{Im}[s_c] \ll 1$. This expression shows that the dynamical susceptibility diverges as the critical points move on to the real-axis, $\mathrm{Im}[s_c]\rightarrow 0$, and the transition becomes sharp.
}

We now apply our method to two open quantum systems which exhibit dynamically intermittency together with a first-order trajectory phase transition.

\begin{figure*}
\includegraphics[width = 0.98\textwidth]{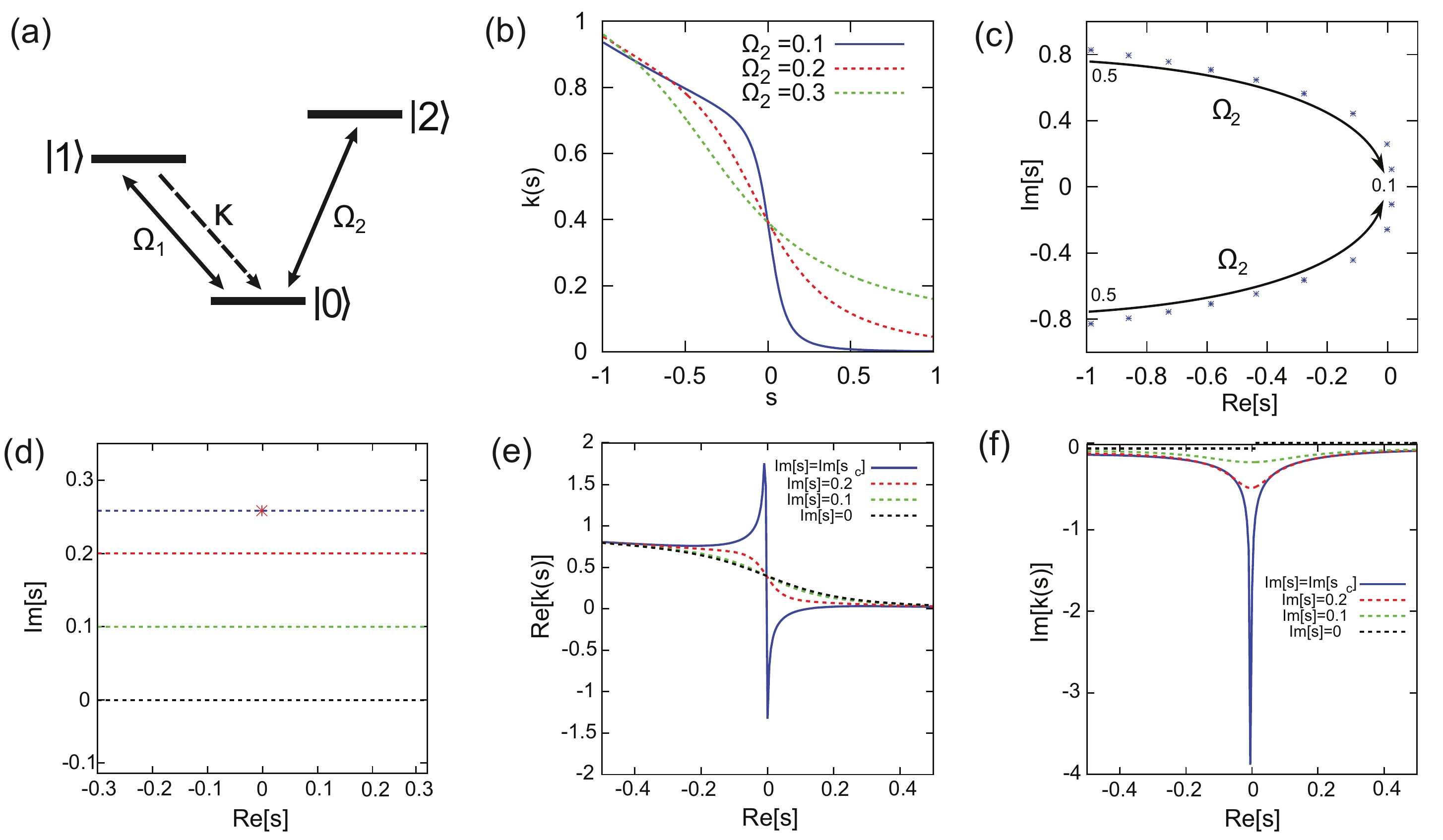}
\caption{(Color online) Driven three-level system. (a) Two lasers with frequencies $\Omega_{1}$ and $\Omega_{2}$ are on resonance with transitions between the ground state $\ket{0}$ and the excited states $\ket{1}$ and $\ket{2}$. The system decays from the state $\ket{1}$ under photon emission with rate $\kappa$. (b) Average dynamical activity $k(s)$ as a function of the counting field $s$. For $\Omega_{1} = 1$, $\kappa = 4$, and  $\Omega_{1}\gg \Omega_{2}$, the dynamical activity exhibits a smooth first-order transition around $s=0$ between an active and an inactive phase. (c) Phase transition points in the complex plane of the counting field \revision{(in the long-time limit)}. As $\Omega_{2}$ is decreased, the phase transition points approach the origin. (d) We now fix $\Omega_{2} = 0.15$ and consider the dynamical activity along the horizontal lines in the complex plane of the counting field. A phase transition point is marked with an asterisk. The real and imaginary parts of the dynamical activity calculated along the horizontal lines are shown in the following panels. (e) The real-part of the dynamical $\text{R}\text{e}[k(s)]$ becomes sharper as the critical point is approached and eventually develops a discontinuity. (f) The imaginary part of the dynamical activity $\text{I}\text{m}[k(s)]$ diverges as the phase transition point is approached.}
\label{fig:fig2}
\end{figure*}

\section{Driven Three-Level System}
\label{sec:3level}

We first consider a three-level quantum system driven by two resonant lasers as shown in Fig.~\ref{fig:fig2}a. The lasers with frequencies $\Omega_{1}$ and $\Omega_{2}$ are on resonance with transitions between the ground state $\ket{0}$ and two excited states of the system denoted as $\ket{1}$ and $\ket{2}$, respectively. The laser-driven oscillations between the ground and excited states are described by the Hamiltonian
\begin{equation}
\hat{H} = \sum_{j=1}^2\Omega_{j}(\hat{a}_j^\dagger+\hat{a}_j),
\end{equation}
where $\hat{a}_j=\ket{j}\bra{0}$ and $\hat{a}_j^\dagger=\ket{0}\bra{j}$, $j=1,2$. In addition, the system is assumed to decay from the state $\ket{1}$ to the ground state $\ket{0}$ with rate $\kappa$ under the emission of a photon. This can be described by the single jump operator
\begin{equation}
\hat{L}_1 = \sqrt{\kappa} \hat{a}_1,
\end{equation}
thus fixing $j=1$ in Eqs.~(\ref{eq:master}) and (\ref{eq:sliouv}). In the following we take the number of emitted photons as the dynamical activity that characterizes the dynamical trajectories.

\begin{figure*}
\includegraphics[width = 0.99\textwidth]{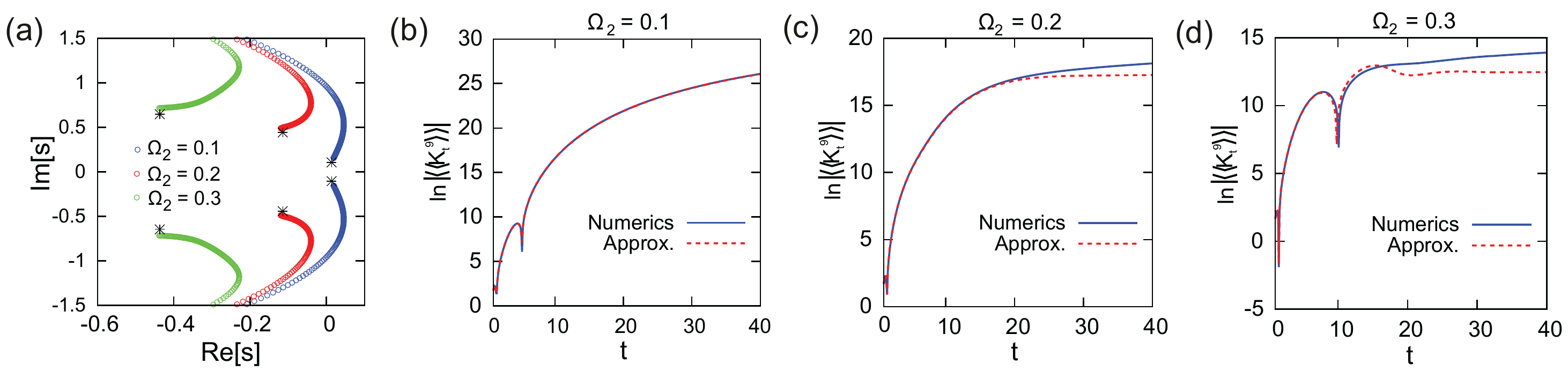}
\caption{(Color online) Lee-Yang zeros and high-order cumulants. (a) Dynamical Lee-Yang zeros extracted from the high-order cumulants  of the dynamical activity.  The leading pair of dynamical Lee-Yang zeros  converge with time to the phase transition points (marked with asterisks) in the complex plane of the counting field. The parameters are $\Omega_{1} = 1$, $\kappa = 4$ and $\Omega_2=$ $0.1$, $0.2$, and $0.3$. (b,c,d) Numerically exact results for the cumulants of order $n=9$ (full lines) as functions of time together with the approximation given by Eq.~(\ref{eq:approx}) (dashed lines) based on the extracted pair of dynamical Lee-Yang zeros in panel (a).}
\label{fig:fig3}
\end{figure*}

The dimensionality of the system at hand is so small that we can directly diagonalize the $s$-dependent Liouvillian to obtain the dynamical free energy $\theta(s)$ together with other $s$-dependent quantities. This in turn allows us to test our method against exact results before moving on to a more complex problem, where we can no longer rely on exact diagonalization.

In Fig.~\ref{fig:fig2}b, we show results for the average dynamical activity, Eq.~(\ref{eq:ave_dyn_act}), as a function of the counting field~$s$. We focus on a parameter regime, where the time scales of the transitions $|0\rangle \rightleftarrows |1\rangle$ are much shorter than those of the transitions $|0\rangle \rightleftarrows |2\rangle$, i.~e., $\Omega_{1},\kappa \gg \Omega_{2}$. In this case, the system displays intermittency in its photon emission trajectories~\cite{Plenio1998,Barkai2004,Garrahan2010}, which manifests itself as a crossover in the dynamical activity as the counting field is varied, see Fig.~\ref{fig:fig2}b. For $s \lesssim 0$, the dynamics is dominated by the highly active transition $\ket{1}\rightarrow \ket{0}$, whereas for $s \gtrsim 0 $, the dynamics is dominated by long periods in the state $\ket{2}$. The intermittency around $s = 0$ may then be viewed as the result of dynamical phase
coexistence between these active and inactive phases.

\revision{For any finite $\Omega_{2}$, the activity is smooth and continuous as seen in Fig.~\ref{fig:fig2}b. As discussed in Sec.~\ref{sec:LYZ}, the smooth crossover is indicative of an avoided crossing between the two largest eigenvalues for real values of the counting field. In Fig.~\ref{fig:fig2}c, we show the complex values of the counting field for which the two largest eigenvalues cross. These critical points move toward the origin as $\Omega_{2}$ approaches zero, in agreement with the transition in Fig.~\ref{fig:fig2}b becoming sharper in this limit. In the lower panels of Fig.~\ref{fig:fig2} we consider the dynamical activity $k(s)$ for complex values of the counting field. We fix the imaginary part of the counting field and vary its real part along the horizontal lines in the complex plane indicated in Fig.~\ref{fig:fig2}d.  The corresponding real and imaginary parts of the dynamical activity are shown in Figs.~\ref{fig:fig2}e and f, respectively.}  When $\text{I}\text{m}[s] = 0$, we see the dynamical crossover in $\text{R}\text{e}[k(s)]$ around $s = 0$, consistent with the dynamical intermittency. Increasing $\text{I}\text{m}[s]$ towards $\text{I}\text{m}[s_c]$, the crossover becomes sharper, until a discontinuity appears at the trajectory critical point.  Similarly, for the imaginary part of the dynamical activity, a sharp minima appears around the critical point. These results corroborate the idea that the dynamical crossover and intermittency are essentially due to a trajectory phase transition in the complex plane.  The remnants of the discontinuity in $k(s)$ at the (complex) critical point appear as a smooth crossover on the real-axis.

Having understood the trajectory phases of the model, we now apply our method based on high-order cumulants of the dynamical activity and dynamical Lee-Yang zeros. The high-order cumulants at finite times are calculated numerically exact using a method introduced in Ref.~\cite{Kambly2011} and also described in Appendix~\ref{sec:Fact}. We recall that the extraction of the dynamical Lee-Yang zeros from the high-order cumulants relies only on the real physical dynamics taking place without the counting field ($s = 0$), which in principle is directly experimentally accessible.

In Fig.~\ref{fig:fig3}a, we show the motion of the leading pair of dynamical Lee-Yang zeros for three different values of $\Omega_2$. The leading pair of dynamical Lee-Yang zeros, $s_0(t)$ and $s_0^*(t)$ are extracted from the time-dependent cumulants of order $n = 6,7,8,9$ by solving Eq.~\eqref{eq:MatEq}. \revisiontwo{The dynamical Lee-Yang zeros are extracted from the cumulants at short times, before they become linear in time, see Figs.~\ref{fig:fig3}b-d.} Figure~\ref{fig:fig3}a shows how the leading pair of Lee-Yang zeros converge to the trajectory critical points with increasing time. \revision{Importantly, already from the leading pair of Lee-Yang zeros we can locate the trajectory critical points and we do not have wait until \emph{all} dynamical Lee-Yang zeros have reached the critical points in the long-time limit. Moreover, as $\Omega_{2}$ is decreased the Lee-Yang zeros converge to points closer to the origin in accordance with the results based on exact diagonalization, see Fig.~\ref{fig:fig2}c.

Figure~\ref{fig:fig3}a reveals a slight difference between the leading pair of Lee-Yang zeros and the actual critical points. The accuracy of our method can be improved by using higher cumulants for which the approximation in Eq.~(\ref{eq:approx}) becomes better. In an experiment, however, the difficulty of precisely measuring cumulants increases with the order of the cumulants. As such, our methods comes with a trade-off between, on the one hand, accurately locating the transitions points and, on the other hand, overcoming the difficulty of measuring high cumulants.}

To further check the extraction of the leading pair of dynamical Lee-Yang zeros, we plug the extracted zeros back into Eq.~(\ref{eq:approx}) and show in Figs.~\ref{fig:fig3}b-d the approximation of the high-order cumulants together with the exact results (here for the cumulant of order $n=9$). The method works best at short times before the contribution from sub-leading Lee-Yang zeros becomes important. We conclude that from a direct measurement of the statistics of photon emission (with $s=0$), one may deduce the position of the trajectory critical points, even if they occur at non-zero values of the counting field.

We now move on to a many-body problem, where a direct numerical solution is not readily available.

\begin{figure*}
\includegraphics[width = 0.98\textwidth]{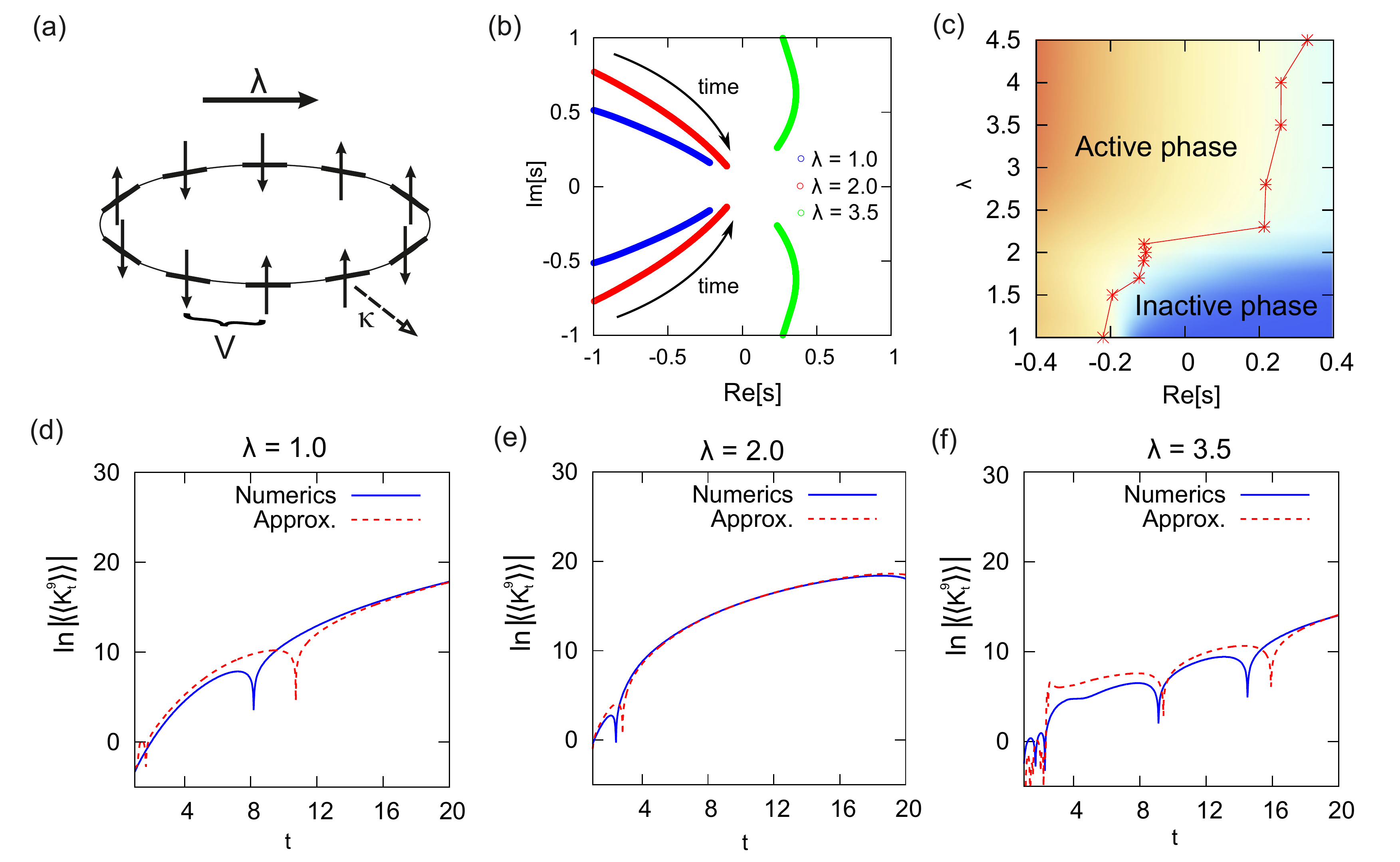}
\caption{(Color online) Dissipative Ising model. (a) The system consists of $N$ spins in the transverse field $\lambda$ and nearest-neighbor interaction $V$. The spins decay with rate $\kappa$. (b)  Motion of the leading pair of dynamical Lee-Yang zeros for different values of the transverse field $\lambda$. The other parameters are $\kappa = 0.1$, $V = 10$, and  $N = 12$.  The Lee-Yang zeros converge to the complex values of the counting field, where the dynamical free energy displays singular features corresponding to a phase transition. (c) Phase diagram showing the real-part of the trajectory transition points for different values of $\lambda$. The line separates the active and the inactive phases. \revision{We also show a density plot of $k(s)$ computed by exact diagonalization of $\mathbb{L}(s)$ for systems of size $N=4$.} (d,e,f) Numerically exact results for the cumulants of order $n=9$ (full lines) together with the approximation given by Eq.~(\ref{eq:approx}) (dashed lines) based on the extracted pair of dynamical Lee-Yang zeros in panel (b).}
\label{fig:fig4}
\end{figure*}

\section{Dissipative Ising Model}
\label{sec:Ising}
As our second application, we consider the dissipative Ising model consisting of $N$ spin-$1/2$ particles arranged on a one-dimensional ring as depicted in Fig.~\ref{fig:fig4}a. The spins evolve coherently under a transverse magnetic field and are described by the Hamiltonian~\cite{Sachdev2011,Ates2012}
\begin{equation}
\hat{H} = \lambda \sum_{j=1}^N \hat{S}_{j}^{(x)} + V \sum_{j=1}^N \hat{S}_{j}^{(z)}\hat{S}_{j+1}^{(z)},
\label{eq:Ising_H}
\end{equation}
where $\hat{S}_j^{(x)}$ and $\hat{S}_j^{(z)}$ are the Pauli spin-$1/2$ matrices for the spin on site $j$. The strength of the transverse field is denoted as $\lambda$ and $V$ is the coupling between neighboring spins. The periodic boundary conditions imply that $\hat{S}_{N+1}^{(z)}$ should be substituted by $\hat{S}_{1}^{(z)}$ in Eq.~(\ref{eq:Ising_H}). Dissipation is provided by a zero temperature bath, causing decay of the excited spin states under photon emission with rate $\kappa$ as described by the Lindblad operators
\begin{equation}
\hat{L}_{j} = \sqrt{\kappa}\hat{S}_{j}^{(-)}.
\end{equation}
Here, the spin lowering operator on site $j$ is denoted as $\hat{S}_{j}^{(-)}$. We again take the number of emitted photons as the dynamical activity.

In recent work, a dynamical phase diagram was constructed within the $s$-ensemble formalism using both numerical methods and mean-field theory~\cite{Ates2012,Lee2011,Lee2012}. It was shown that depending on the ratios $\lambda/\kappa$ and $V/\kappa$, the system is either in an active phase or an inactive phase.  Fixing the ratio $V/\kappa$, the mean-field theory also predicted that an intermittent phase can exist between the two phases for a range of $\lambda$ values.  However, numerically it appeared that this intermittent phase collapses to a single transition point $\lambda_c$ in the large system-size limit, $N\rightarrow\infty$.  At this point, the system will be at coexistence between the active and inactive states and will show intermittency, marked by a first order trajectory transition at $s = 0$.  Furthermore, in the region around $\lambda_c$, but firmly in the active or inactive phases, one would expect the full distribution of quantum jumps to
possess some features of the other phase due to the proximity of the transition point.

This behavior may be probed by tuning the counting field $s$ and should manifest itself as trajectory transition points at $s \neq 0$. To confirm this prediction, we calculate the high-order cumulants of the dynamical activity and extract the leading pair of Lee-Yang zeros. Doing this for different values of the transverse field $\lambda$, we can construct a trajectory phase diagram for the dissipative Ising model. We tune the system across a range of transverse fields, where the system changes from being inactive with the stationary state being ferromagnetic in nature (all spins pointing down) to being highly active with the stationary state being paramagnetic. We consider a finite-size chain, where the crossover between the two phase is smeared out, but we expect that the transition becomes sharp in the large system-size limit, $N\rightarrow\infty$.

To find the dynamical Lee-Yang zeros, we prepare the system close to its equilibrium state in each parameter regime and then evaluate numerically the high-order cumulants using the method described in Appendix~\ref{sec:Fact}. From the cumulants of order $n =6,7,8,9$ we then extract the leading dynamical Lee-Yang zeros as functions of time using Eq.~\eqref{eq:MatEq}. The results for three different value of the transverse field $\lambda$ are shown in Fig.~\ref{fig:fig4}b. Below a certain threshold value of the transverse field, the dynamical Lee-Yang zeros converge to a complex pair of phase transition points with negative real-parts, showing that the system at $s=0$ is mostly in the inactive phase. At the transverse field is increased, the system eventually crosses over into the active phase with corresponding phase transition points that have positive real-parts. This is similar to the results for the driven three-level system, where the dynamical phase coexistence is characterized by critical point points away from the origin.

We can now construct a trajectory phase diagram for the dissipative Ising model. In Fig.~\ref{fig:fig4}c, we show the real-part of the critical points $s_c$ and $s_c^*$ for different values of~$\lambda$. We see that the system shows a rather sharp transition within a narrow band of transverse fields even for this finite-size system, and our results (for $s=0$) are consistent with the numerical calculations as well as the mean-field theory presented in Ref.~\cite{Ates2012}. Finally, to check the consistency of our results, we insert the extracted Lee-Yang zeros back into Eq.~\eqref{eq:approx} and compare the result with the numerically calculated cumulants. This comparison is made in Fig.~\ref{fig:fig4}d-f, showing the extraction of the leading dynamical Lee-Yang zeros works well for the periods of time that we consider here.

\section{Conclusions}
\label{sec:Conc}

We have used high-order cumulants of dynamical observables to investigate the dynamical Lee-Yang zeros of generating functions in open quantum systems. With our recently proposed method, we have investigated the photon emission processes for a driven quantum three-level system and for the dissipative quantum Ising chain. These systems are of particular interest as they exhibit dynamical intermittency accompanied by a first-order trajectory phase transition. The phase transitions are manifested as crossovers in the dynamical activity as an external counting field is varied. However, as we have shown, the existence and location of the phase transitions can also be inferred from the dynamical Lee-Yang zeros as they with time move towards the phase transition points in the complex plane of the counting field. Importantly, the Lee-Yang zeros can be extracted from the high-order cumulants of the dynamical activity corresponding to the real physical dynamics taking place without the counting field. As such, our method offers the possibility to detect trajectory phase transitions in open quantum systems from the finite-time behaviour of measurable quantities.
The approach is quite general and may be extended to other observables and associated ensembles of trajectories such as quadrature trajectories. In future work, it would be interesting to understand if the nature of a phase transition (first-order or continuous) can be understood from the way the dynamical Lee-Yang zeros accumulate in the long-time limit. This may require that not only the leading pair of Lee-Yang zeros is extracted from the high-order cumulants, but also sub-leading pairs of zeros might be needed.

\section{acknowledgements}
The work is supported by the NCCR QSIT, by the Swiss NSF, by EPSRC Grant no. EP/I017828/1, and by Leverhulme Trust grant no.~F/00114/BG.

\appendix

\section{Calculation of high-order cumulants}
\label{sec:Fact}

To calculate high-order cumulants at finite times we follow the method developed in Ref.~\cite{Kambly2011}. We first define a generating function of factorial moments
\begin{equation}
Z_{F}(s,t) = Z(s,t)|_{e^{-s}\rightarrow s+1}=\sum_{K_t}P(K_t,t)(s+1)^{K_t},
\end{equation}
obtained by replacing $e^{-s}$ by $s+1$ in Eq.~(\ref{eq:MGFmas}).  The factorial moments are then given by its derivatives with respect to $s$ evaluated at $s=0$
\begin{equation}
\langle K_t^{n} \rangle_{F} = \partial^{n}_{s}Z_{F}(s,t)|_{s\rightarrow0}= \langle K_t(K_t-1)\ldots(K_t-n+1)\rangle.
\end{equation}
Similarly to the ordinary moments and cumulants, the generating function of the factorial cumulants is
\begin{equation}
\Theta_{F}(s,t) = \ln Z_F(s,t)
\end{equation}
from which the factorial cumulants follow as
\begin{equation}
\llangle K_t^{n}\rrangle_F =
(-1)^{n}\partial^{n}_{s}\Theta_F(s,t)|_{s\rightarrow0}.
\end{equation}
The ordinary cumulants can be expressed in terms of the factorial cumulants via the relation
\begin{equation}
\label{eq:Convert}
\llangle K_t^{m} \rrangle = \sum^{m}_{j=1} \mathrm{st}(m,j) \llangle K_t^{j} \rrangle_{F},
\end{equation}
with $ \mathrm{st}(m,j)$ being Stirling numbers of the second kind.

The advantage of the factorial moments and cumulants becomes clear when considering their dynamical equations. Replacing $e^{-s}$ by $s+1$ in Eq.~(\ref{eq:sliouv}) we find
\begin{equation}
\frac{d}{dt}\hat{\rho}_F(s,t)= \mathcal{L}\hat{\rho}(s,t)+ s\sum_{j} \hat{L}_{j}\hat{\rho}_F(s,t) \hat{L}^{\dagger}_{j},
\end{equation}
having defined
\begin{equation}
\hat{\rho}_F(s,t) = \sum_{K_t}s^{K_t} \hat{\rho}(K_t,t).
\end{equation}
Differentiating the dynamical equation $n$ times with respect to $s$ at $s=0$ we find a hierarchy of coupled equations reading
\begin{equation}
\frac{d}{dt}\hat{\rho}^{(n)}_F(t) = \mathcal{L}\hat{\rho}_F^{(n)}(t)+ n\sum_{j} \hat{L}_{j}\hat{\rho}_F^{(n-1)}(t) \hat{L}^{\dagger}_{j},
\label{eq:hier}
\end{equation}
where
\begin{equation}
\hat{\rho}^{(n)}_F(t)=\partial_s^n\hat{\rho}_F(s,t)|_{s=0}.
\end{equation}
Taking the trace of these density matrices, we find the factorial moments, i.~e.,
\begin{equation}
\langle K_t^{n} \rangle_{F}=\mathrm{Tr}[\hat{\rho}^{(n)}_F(t)]
\end{equation}
from which the factorial and ordinary cumulants follow. We note that Eq.~(\ref{eq:hier}) for the factorial moments only couples $\hat{\rho}^{(n)}_F(t)$ to $\hat{\rho}^{(n-1)}_F(t)$, which is much simpler than the corresponding equations of motion for the ordinary moments.

For small system sizes, like the driven three-level system, the factorial cumulants up to order $N$ can be found using exact diagonalization. To this end, we introduce the vector
\begin{equation}
\mathbf{g} =
\left[|\hat{\rho}(t)\rangle\!\rangle,|\hat{\rho}_F^{(1)}(t)\rangle\!\rangle,\ldots,|\hat{\rho}_F^{(N)}(t)\rangle\!\rangle\right]^{T},
\end{equation}
where $\ket{\hat{\rho}_F^{(n)}(t)}\!\rangle$ is the vector representation of $\hat{\rho}^{(n)}_F(t)$ with $n=1,\ldots,N$. From Eq.~(\ref{eq:hier}) we then obtain the equation
\begin{equation}
\frac{d}{dt}\mathbf{g}  = \mathbf{L}\mathbf{g} ,
\end{equation}
where the matrix $\mathbf{L}$ reads
\begin{equation}
\mathbf{L} = \left[\begin{array}{ccccc}
						\mathbb{L}  &0  &0 &0 &0 \\
						\mathbb{L}'&\mathbb{L} &0 &0 &0 \\
						0  &2\mathbb{L}' &\mathbb{L} &0 &0 \\
						\vdots &\vdots &~ &\ddots & ~ \\
						0 & 0 & 0 & N\mathbb{L}' & \mathbb{L} \\
\end{array}\right]
\label{eq:bigmat}
\end{equation}
with $\mathbb{L}=\mathbb{L}(s=0)$ and $\mathbb{L}'$ being the matrix representation of the quantum jump operators $\sum_{j} L_{j}\bullet L_{j}^{\dagger}$. For the driven three-level system, we can diagonalize the matrix in Eq.~(\ref{eq:bigmat}) and thereby evaluate the time-dependence of $\mathbf{g}$ to obtain the cumulants. For the dissipative Ising model, the resulting matrix is too large to diagonalize, and instead we solve Eq.~(\ref{eq:hier}) using a third-order Runge-Kutta algorithm.

\end{document}